\documentclass[sigconf, nonacm]{acmart}

\AtBeginDocument{%
  }

\begin{document}
\vspace{-0.5in}

\title{Towards Knitted Textile Electromechanical Systems}

\author{John Martins, Abigail Hou, Rishit Garg, Wenchi Liu\footnotemark[1], Brandon Tendilla\footnotemark[1], Noah Tannas\footnotemark[1], Ben Kim, Alejandro Goldstein, Michael Miller, Elizabeth McLaughlin, Nivedita Arora\footnotemark[2]
}

\affiliation[]{
    \institution{\textsuperscript{1}VAK Embodied Systems Lab, Northwestern University}
    \city{Evanston}
    \state{Illinois}
    \country{USA}
}

\renewcommand{\shortauthors}{Martins et al.}

\begin{abstract}
For physical AI systems to dexterously manipulate objects and safely interact with people, they need high-density tactile sensing distributed over large, deformable surfaces, much like human skin. Machine knitting is a promising technique to develop substrates for such tactile sensing; it is additive and scalable, and it produces stretchable, mechanically programmable structures. Achieving skin-level tactile resolution in knitted textiles, however, requires conductive yarns with an insulating sheath that prevent electrical shorting between adjacent tactile pixels. The insulation must remain conformal and intact under the tight loop curvature and high flexural strain imposed by the knitting process, a requirement that existing commercially-available, insulated conductive yarns do not meet.

In this work, we present a yarn dip-coating process to create insulated conductive yarn optimized for machine knitting. Process parameters for coating silver-plated nylon yarns with thermoplastic polyurethane (TPU) are guided by an adjusted dip-coating fluid dynamics model and linked to yarn-level electrical and mechanical characteristics, showing compatibility with machine knitting. Furthermore, we see minimal deviation in these characteristics after machine knitting and washing.

Knitted capacitive pressure sensors with multi-layered structures are created to demonstrate e-textile applications. Our process demonstrates a viable manufacturing technique to produce functional yarns that paves the way for scalable, high-density tactile sensing embedded directly in textile structures.
\end{abstract}


\begin{teaserfigure}
 \includegraphics[width=\textwidth]{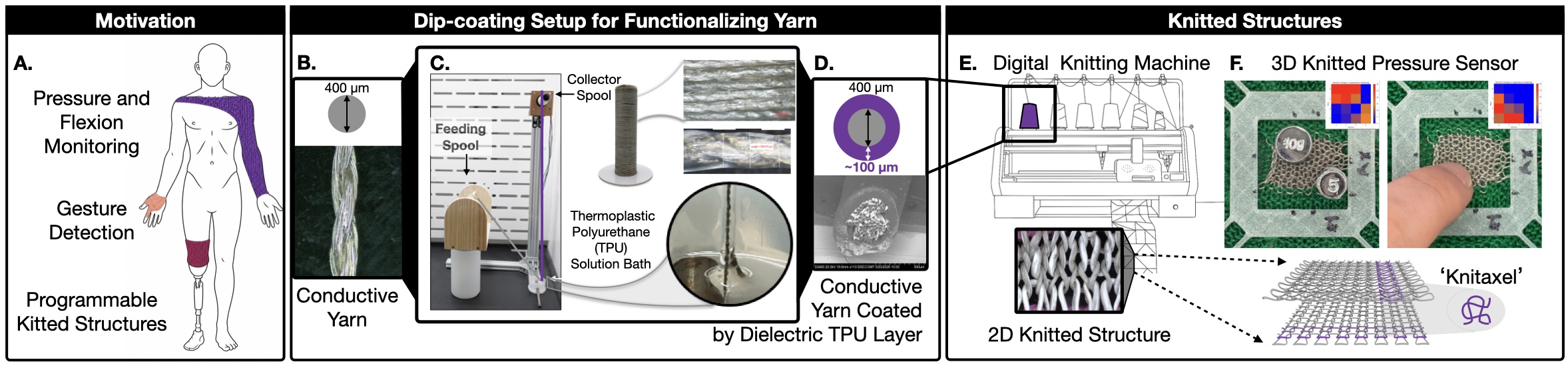}
  \caption{Dip-coating enables mechanically compliant, insulated coatings for machine knittable, functionalized yarns applied in e-textiles. (A) Wearable knitted interfaces for gesture and pressure sensing. (B) Pre-coated silver-plated nylon yarn. (C) Dip-Coating process. (D) Diagram and cross-section of custom coated thermoplastic insulated silver-plated nylon yarn. (E) Programmable Digital Kniterate Machine. (F)  Prototype of two-layer knitted capacitive sensing pressure array}
  \Description{}
  \label{fig:tems_figure}
\end{teaserfigure}

\maketitle

\footnotetext[1]{These authors contributed equally to this research}
\footnotetext[2]{Corresponding Author: nivedita@northwestern.edu}

\pagestyle{plain}

\section{Introduction}
Machine knitting is receiving increasing attention as an additive fabrication method for e-textiles and wearable sensors because it enables scalable production of inherently stretchable and mechanically programmable garments \cite{singal_programming_2024, pasquier_multi-level_2025}. Knitted e-textiles \cite{Takahashi_Knittable_2026, aigner_loopsense_2024, Pointner_Preindl_Mlakar_Aigner_Haberfellner_Haller_2022, McDonald_Vallett_Solovey_Dion_Shokoufandeh_2020} offer stretchability and dynamic geometries that woven, braided, and yarn-inlaid fabrics \cite{sun_weaving_2020, batra_texteng_2025, luo_learning_2021} lack, making them well suited for wearable interfaces (Figure \ref{fig:tems_figure}.A.) \cite{Luo_Zhu_Wu_Honnet_Mueller_Matusik_2023, kim_knitdermis_2021, fan_machine-knitted_2020, billard_good_2024, liu_wearability_2016}. Progress in human-computer interaction, health monitoring, and robotic tactile sensing relies on developing hardware that brings sensing and feedback towards the fidelity of skin mechanoreceptors \cite{billard_good_2024}. This has yet to be shown in conformable, knitted e-textiles. Reaching this density requires insulated conductive yarns compatible with machine knitting and able to withstand tight, looped geometry without shorting or compromising mechanical compliance. Functionalized yarns have not been engineered specifically for machine knitting and the link between yarn materials, fabrication processes, mechanical properties, and sensing performance remains largely unquantified or stops short of sensing characterization \cite{alagirusamy_coating_2013, singal_programming_2024}. This motivates the need for creating insulated, functionalized yarns that are optimized for machine-knitting.

This work introduces a simple and accessible yarn fabrication technique whereby silver-plated nylon conductive yarns are dip-coated with thermoplastic polyurethane (TPU) insulation. A base fluid dynamics framework guides initial dip-coating parameters towards optimal coating conditions that minimize yarn stiffness while maintaining electrical integrity.

\begin{figure}[htbp]
    \centering
    \includegraphics[width=\linewidth,height=0.9\textheight,keepaspectratio]{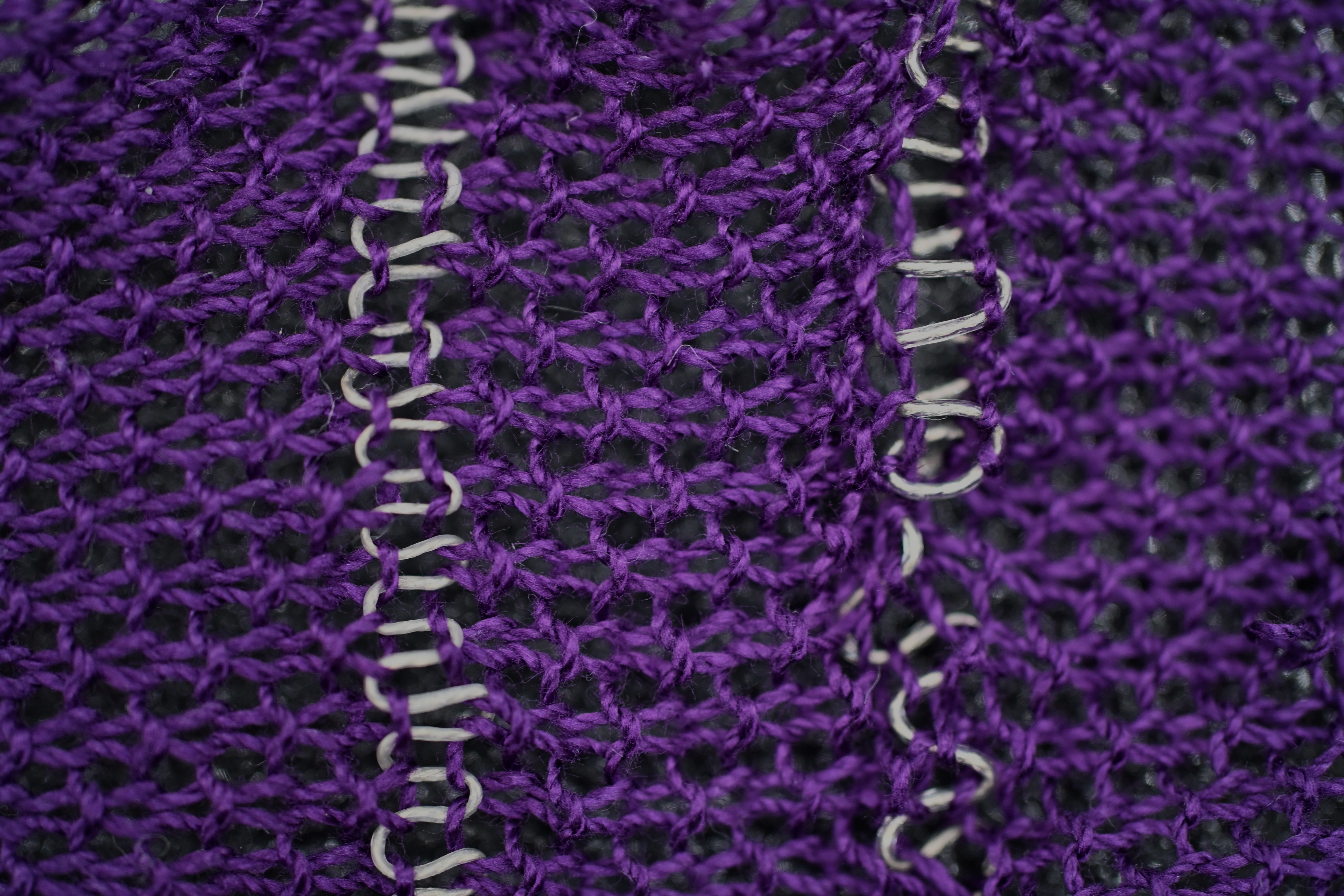}
    \caption{Side by side comparison of our custom coated yarns (left) and commercially available TPU-coated yarn (right) knitted using the seed stitch pattern. They are similar gauge yarn and use the exact same core conductive yarn, but the commercially available coated option displays unwinding from the looped structure due to high flexural stiffness. This patch was imaged within 5 minutes of being machine knitted and is under no pretension or strain.}
    \label{fig:shieldex_custom_comparison}
\end{figure}

The electrical and mechanical properties of the dip-coated yarns are characterized and compared to commercially available TPU-coated yarns of similar gauge that experimentally demonstrated difficulty knitting and frequent unwinding from the knit structure (Figure \ref{fig:shieldex_custom_comparison}). The high flexural stiffness of commercially available insulated yarn limits stitch density and wearability \cite{Gemperle_Kasabach_Stivoric_Bauer_Martin_1998}. Our yarns show comparable electrical insulation that supports the e-textile function and superior mechanical properties compatible with machine knitting. We additionally show that the machine knitting process and washing have negligible effects on electrical and mechanical properties.

Lastly, preliminary knitted applications are created with a digital flatbed knitting machine (Figure \ref{fig:tems_figure}.E.) and demonstrate capacitive pressure sensing capabilities. We discuss future application directions that include multi-model integrated textile sensing through more complex, plated knit geometries.

%Add in the fabrication section below:
%"2-ply Shieldex 235/36" --> with citation \cite{shieldex_235_36_TPU}
%Commercially-available 2-ply Shieldex 235/36 TPU Coated..." --> \cite{shieldex_235_36_TPU} 

%Figures could be used/inspired from our other unreleased preprint
%Figures to add here:
% - dip-coating Apparatus
% - Microscopy of yarns
% - Viscosity Data graph

\section{Dip-Coating Fabrication Process} 

\begin{figure*}[htbp]
    \centering
    \includegraphics[width=\textwidth,height=0.5\textheight,keepaspectratio]{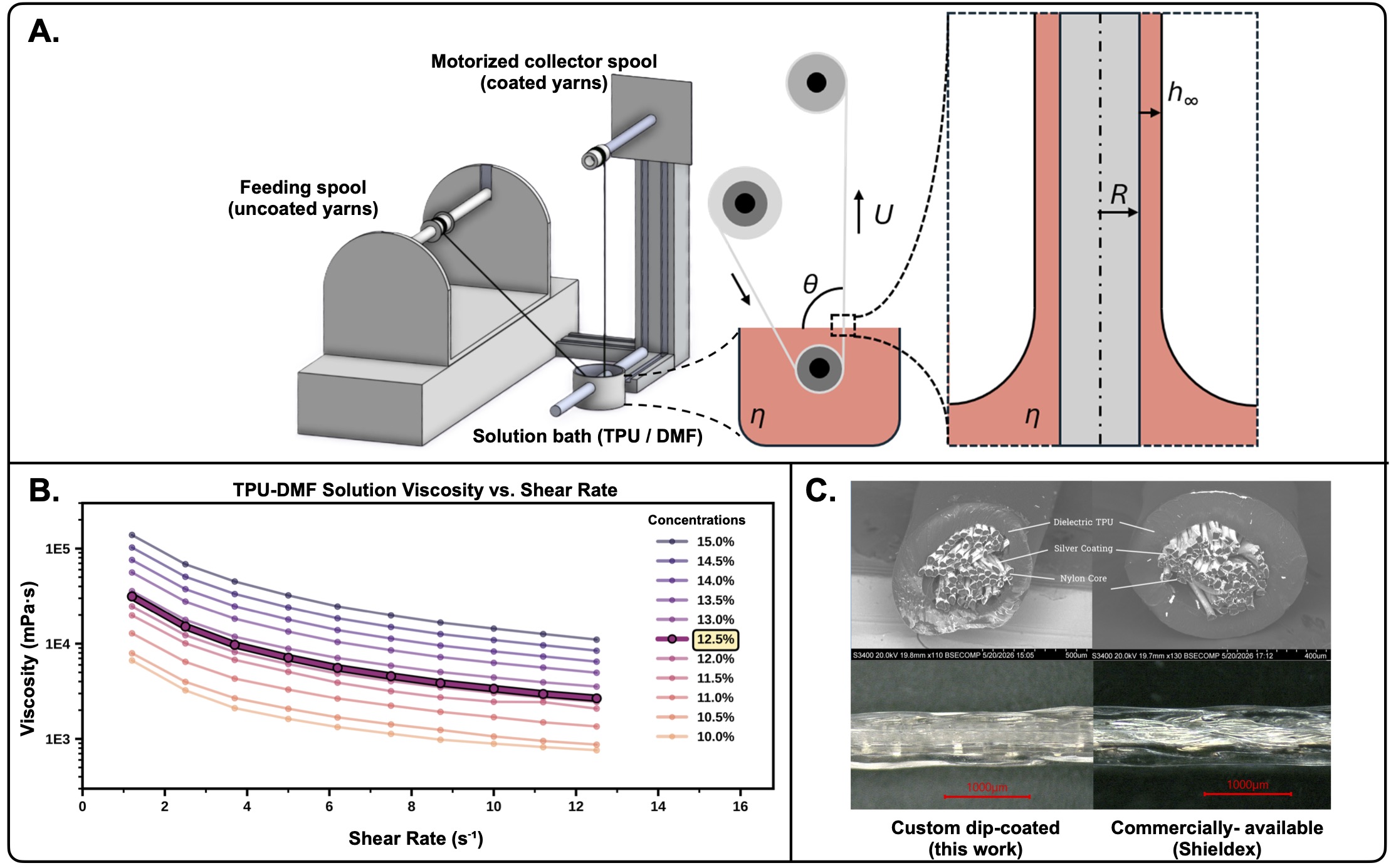}
    \caption{Fabrication of dip-coated yarns. (A) A model of the custom dip-coating process with zoomed in portions on the yarn path through the solution bath and meniscus of fluid entrainment on the yarn. Starting from feeding spool, yarns are pulled through the TPU/DMF solution bath by the motorized collector spool, where they accumulate to dry. Variables shown represent the properties of the solution bath, dip-coat process, and resulting film thickness, where $h_\infty$ is the thickness of the fluid film deposited, $\gamma$ is surface tension,  $\eta$ is the solution viscosity, $R$ is the fiber radius, $U$ is withdrawal speed, and $\theta$ is the withdrawal angle. (B) Shear-thinning viscosities measured for TPU-DMF concentrations between 10\% and 15\%. We highlight 12.5\% as our selected concentration that experimentally produces the most consistent results and theoretically fits our target coating thickness of \textasciitilde 100 $\mu m$. (C) Cross-sectional SEM images(top row) and length-wise optical microscopy images (bottom row) of the knitting-optimized dip-coated yarn(left) and commercial TPU-insulated Shieldex reference(right).}
    \label{fig:dipcoat_viscosity_microscopy}
\end{figure*}

We begin with 2-ply Shieldex 235/36 silver-plated conductive yarns \cite{shieldex_235_36_TPU}, chosen for having a diameter compatible with standard 6-8 gauge knitting machines, low flexural stiffness, and being widely available. To prepare for the dip-coating procedure, conductive yarns are aligned and tensioned between a source feeding spool and a collector spool to ensure uniform wetting and minimize geometric variability along the filament length. The collector spool is connected to a motorized system designed to maintain a constant speed and positioned to ensure a perpendicular withdrawal angle throughout the coating process. Yarns are drawn from the feeding spool and through a solution bath created by dissolving 85A hardness TPU in dimethylformamide (DMF), which evaporates away once coated onto the yarns. Lastly, to ensure complete DMF evaporation, coated yarns are left to dry on the collector spool under slight airflow in a laboratory fume hood for a minimum of 20 hours before use.

From an adjusted Landau-Levich-Derjaguin (LLD) model for thin fibers \cite{zhang_dip_2022}, it is known that in dip-coating, solution viscosity, surface tension, and dip-coat withdrawal speed directly govern film entrainment and serve as physical bases for predicting coating thickness. To rationalize the influence of our solution properties on coating behavior, we systematically varied the TPU wt.\% of our solution across the concentration range 10\% to 15\% while maintaining a constant withdrawal speed of 2.5 mm/s. Experimentally, we determined that altering just the TPU wt.\% led to the most pronounced and controllable changes in yarn coating thickness; surface tension remained constant across all wt.\% solutions tested, thus proving difficult to change without extra solutes, and withdrawal speed adjustments also impacted dynamic viscosity due to the shear-thinning behavior of the TPU/DMF solution. We identify conditions that produced continuous, defect-free TPU insulation layers with suitable bending stiffness for machine knitting. The LLD model relating these parameters with coating thickness for an assumed 90 degree withdrawal angle is shown in Equation \eqref{eq:delta} below:

\begin{equation}
    h_\infty = 1.34\, (\mathrm{\frac{\eta U}{\gamma}})^{2/3} R\label{eq:delta}
\end{equation}

where $h_\infty$ is the thickness of the fluid film deposited ($m$), $\gamma$ is surface tension ($N/m$),  $R$ is the fiber radius ($m$), $U$ is withdrawal speed ($m/s$), and $\eta$ is the solution viscosity ($Pa \cdot s$).

% \begin{figure}[H]
%     \centering
%     \includegraphics[width=\linewidth,height=0.9\textheight,keepaspectratio]{viscosity_B_highlight_12p5.png}
%     \caption{Sweep of viscosity measurements vs. rotary viscometer shear rates for 11 dip-coating solutions varying from 10 - 15 TPU wt.\%. A non-Newtonian, shear-thinning behavior is seen with viscosity decreasing with power law relationship vs. shear rate. The 12.5 TPU wt.\% is highlighted showing our selected solution concentration as it both experimentally produces the most consistent results and theoretically fits our target coating thickness of \textasciitilde 100 $\mu m$}
%     \label{fig:viscosity}
% \end{figure}

We measured surface tension and viscosity using the pendant-drop method \cite{Juza_2019} and a rotary viscometer respectively. Across our experimental sweep of solution TPU wt.\%, we observed a constant surface tension of 34.76 mN/m and a non-Newtonian, shear-thinning viscosity as shown in Figure \ref{fig:dipcoat_viscosity_microscopy} B. It is recognized that additional models and future modifications to the LLD equation are necessary to account for non-Newtonian solution behavior \cite{afanasiev_landau-levich_2007} as well as the complex wound fiber geometries and a non negligible yarn radius to coating thickness ratio.

%  \begin{figure}[htbp]
%     \centering
%     \includegraphics[width=\linewidth,height=0.9\textheight,keepaspectratio]{sem_optical_microscopy.png}
%     \caption{Cross-sectional SEM images(top row) and length-wise optical microscopy images (bottom row) of the knitting-optimized dip-coated yarn(left) and commercial TPU-insulated Shieldex reference(right) showing the conductive filament bundle and surrounding dielectric shell. }
%     \label{fig:microscopy}
% \end{figure}

As such, we evaluated experimental measurements against theoretical prediction to identify an optimal 12.5\% TPU-weighted DMF formulation that allowed for machine knitting-compatible yarn tensile strength, bending stiffness and coating diameter after just 1 pass through the solution bath. Due to difficulty matching shear-dependent viscosity measurements exactly with the shear experienced by yarn coating at the chosen 2.5 mm/s withdrawal speed, future work is geared towards understanding the effective viscosity for the 12.5\% solution. This critical first step in understanding rheological properties of our solution sets up more thorough understanding of coating behavior on threads that can further enable coating with different materials optimized for machine knitting. \cite{Javidi_Pope_Hrymak_2016, Tang_Yan_2017}

Yarn coating thickness and quality was assessed with scanning electron and optical microscopy (Figure \ref{fig:dipcoat_viscosity_microscopy} C) and compared to commercially available, TPU-coated 2-ply Shieldex 235/36 \cite{shieldex_235_36_TPU}. We achieve a resultant coated yarn with a diameter of $\sim630 \mu\mathrm{m}$ compared to the diameter of commercially coated yarns at $\sim700 \mu\mathrm{m}$. Our custom coated yarns also achieve similar evenness and surface morphology as commercial options for the 2 meter samples we are currently making. This work provides a reproducible processing pathway to produce high-quality insulated threads suitable for advanced textile integration.

%Figures to add / improve in this section:
% - Standalone leakage current graph (with same colors)
% - Mechanical data with just tensile + bending

\section{Electrical and Mechanical Characteristics}

We demonstrate that our custom, TPU-coated yarns achieve coating insulation levels on par with commercial TPU coatings while also exhibiting more flexibility and knittability than commercially available options.

\begin{figure}[htbp]
    \centering
    \includegraphics[width=\linewidth,height=0.9\textheight,keepaspectratio]{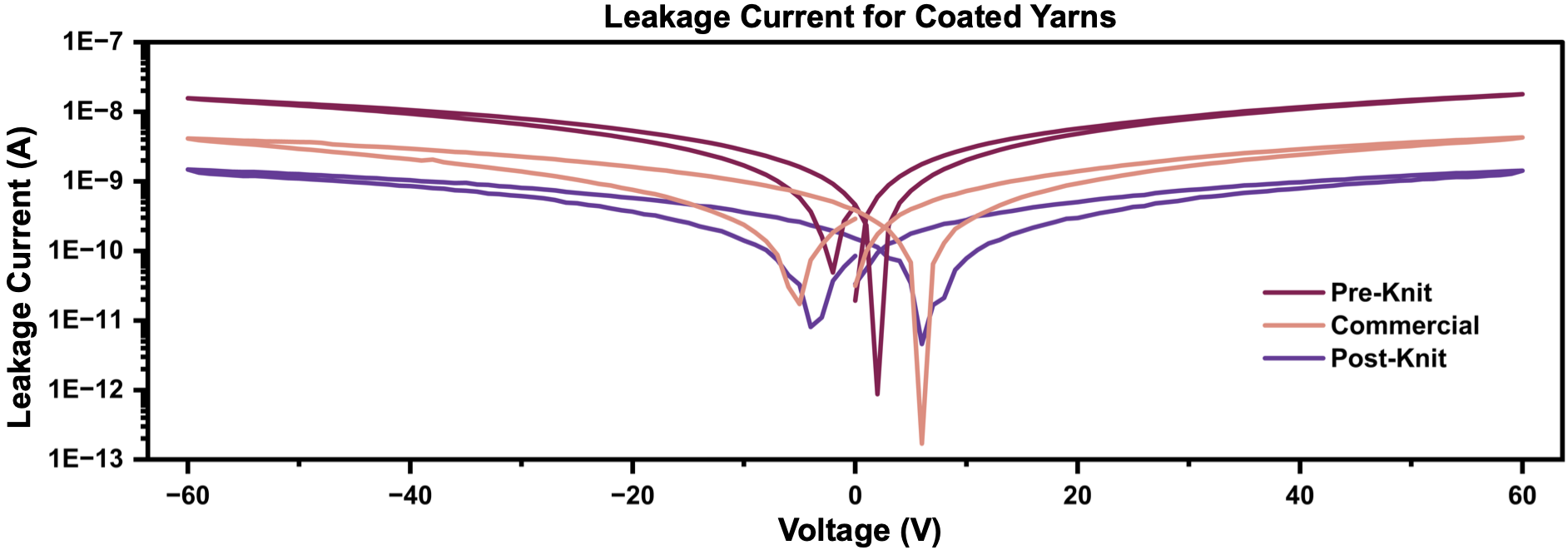}
    \caption{Leakage currents compared between this work's custom dip-coated yarn in a pristine state before knitting ("Pre-knit"), dip-coated yarns after knitting and unraveling ("Post-knit"), and Shieldex TPU coated yarn ("Commercial Coated").}
    \label{fig:leakage}
\end{figure}

Electrical leakage current tests were performed by stretching yarns against a grounded plate and using a probe station to sweep voltages, from -60 to 60 volts, between the yarn core and external plate. Results in Figure \ref{fig:leakage} show that the maximum leakage currents of our dip-coated yarns are within <10 nA of that of commercial coated yarns and show negligible deviation between pre and post knit samples. The similar leakage-current response between the custom and commercial insulated yarns indicates that the dip-coated TPU layer functions as an effective dielectric barrier, even with coating parameters optimized for mechanical compliance required by machine knitting. We see a decrease in leakage current for post-knit dip-coated yarns, however this may be due to slight coating variations across the measured samples. Additional testing with larger sample sizes and more robust testing procedures is underway to more fully validate these findings. 

\begin{figure}[htbp]
    \centering
    \includegraphics[width=\linewidth,height=0.9\textheight,keepaspectratio]{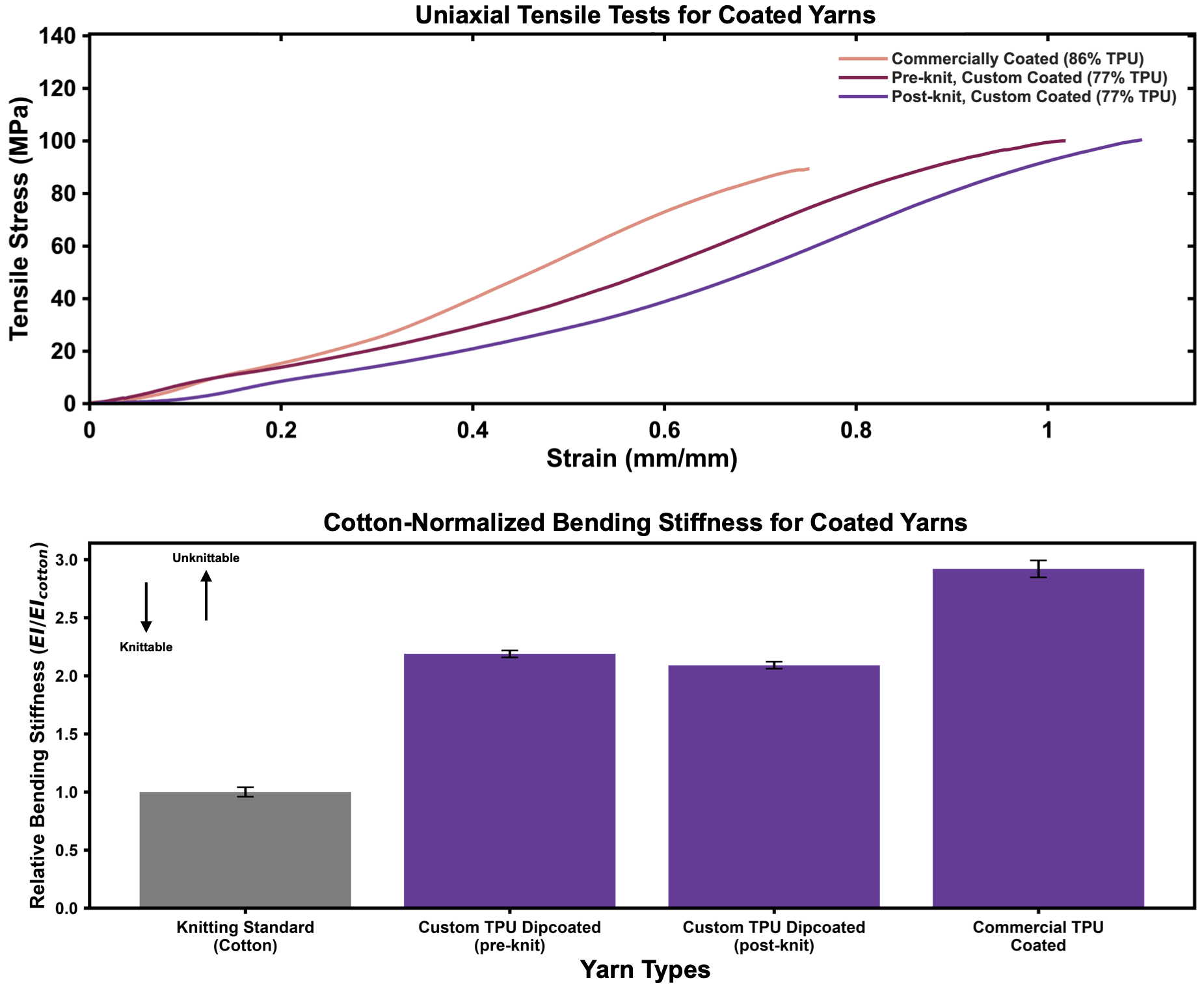}
    \caption{Yarn-level mechanical tensile properties (top) and bending stiffness (bottom) are compared between this work's custom dip-coated yarn in a pristine state before knitting ("Pre-knit"), dip-coated yarns after knitting and unraveling ("Post-knit"), and Shieldex TPU coated yarn ("Commercial Coated"). In the tensile graph, percentages next to yarn labels indicate the percentage of yarn mass that is attributed to TPU coating.}
    \label{fig:mechanical}
\end{figure}

Tensile strength and bending stiffness of yarns were measured through uniaxial tensile testing and the hanging pear loop test \cite{plaut_determining_2020, el_messiry_comparative_2025} respectively.

Our yarns showed superior elasticity and tensile strength during tensile breakage tests as shown in Figure \ref{fig:mechanical}. We observed higher elasticity, yet similar breaking force, in post-knit yarns that we believe is attributed to a softening of the TPU coating and yarn structure from strains experienced during the machine knitting process.

We measured bending stiffness of each yarn type and normalized against a standard, 10/2 count cotton yarn baseline. Results are displayed in Figure \ref{fig:mechanical} and show a bending stiffness of 2.18 and 2.91 times that of cotton for our dip-coated yarn and commercial coated yarns respectively. A slight downward stiffness change was observed for post-knit samples that further indicates a softening of yarn structure from warping experienced during machine knitting. Knittability arrows in the top left corner indicate observed trends in the relationship between bending stiffness in the ranges of the measured yarns and knittability with a 6-8 gauge knitting machine. Future work is aimed at uncovering a more defined knittability range for bending stiffness and other factors, such as surface friction combinations, that may impact machine-knittability \cite{Döonmez_Marmarali_2004}.

%Add in the following Appications section:
%Kniterate® \cite{kniterate_machine} 
\section{Current Applications}

\begin{figure*}[htbp]
    \centering
    \includegraphics[width=\textwidth,height=0.9\textheight,keepaspectratio]{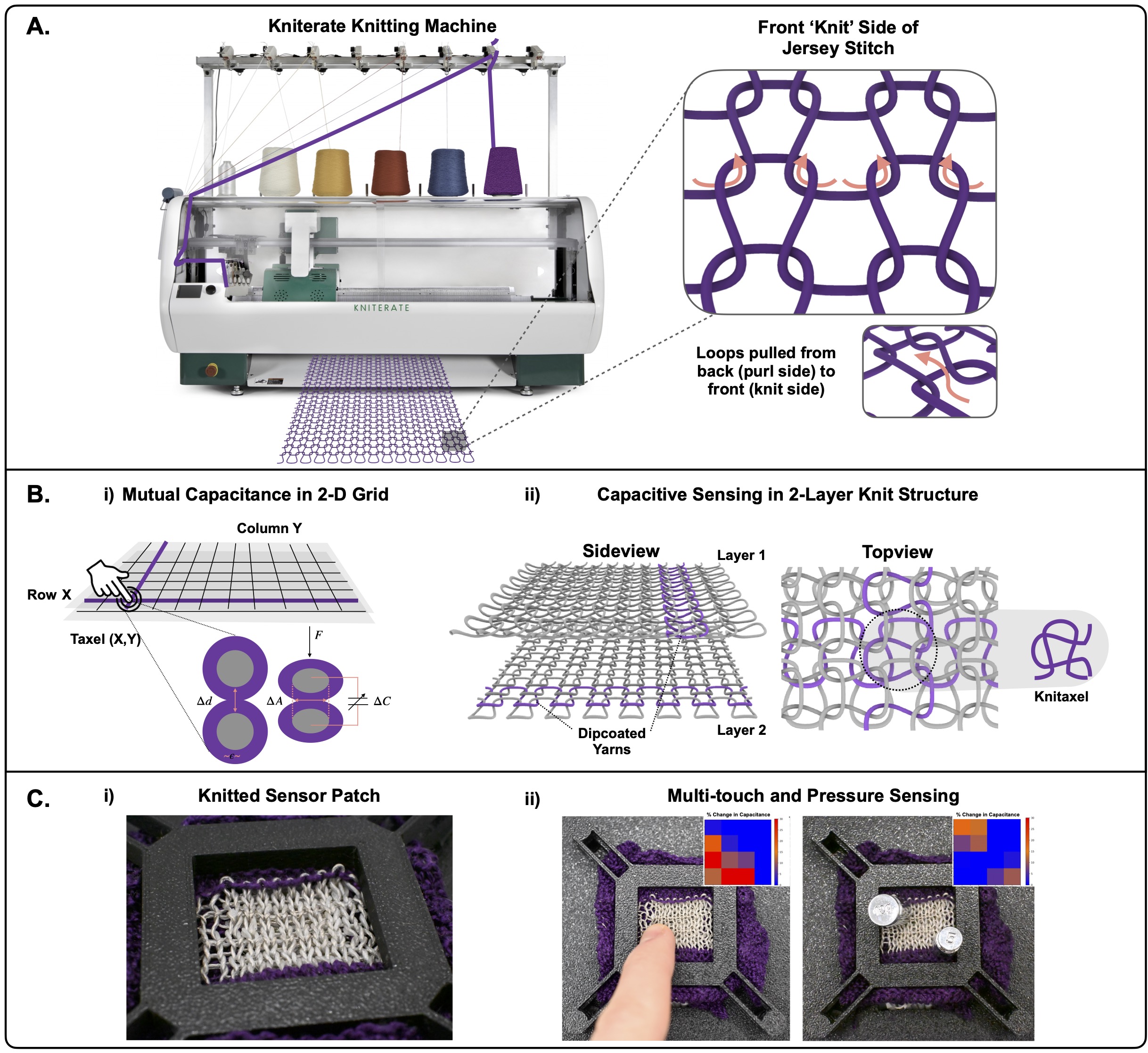}
    \caption{Breakdown of knitting a capacitive sensing patch and demonstration of tactile sensing capabilities. (A) Using the Kniterate knitting machines, yarns form an inter-looped structure such as the example of Jersey knit shown. (B) Mutual capacitance grids (i) allow for 2-D localization and pressure mapping based on force-induced capacitive change. When 2 layers of coated conductive yarns are placed perpendicularly (ii), the fabric itself becomes a capacitive sensing grid of 'Knitaxels'. (C) A sample sensor grid (i) shows touch localization and multi-touch distinguishability of different pressures (ii).}
    \label{fig:application}
\end{figure*}

Knitted applications were performed using a Kniterate® digital knitting machine \cite{kniterate_machine}, a 6-8 gauge, double bed knitting machine (Figure \ref{fig:application} A). We evaluate the performance of our custom dip-coated yarns by knitting them to create high-density capacitive tactile sensors. The sensors are created by knitting two dense patches of insulated conductive yarns using the jersey stitch type and overlaying them perpendicularly to create a capacitive grid of intersecting rows and columns. A microcontroller scans through each combination of row and column, measuring the mutual capacitance between the chosen row and column using an FDC2214 capacitance-to-digital converter \cite{fdc2214}. Location of pressure is determined by identifying the row-column intersection, coined "knitaxel", from which a touch-induced capacitance change is recorded over time (Figure \ref{fig:application} B).

\begin{figure}[htbp]
    \centering
    \includegraphics[width=\linewidth,height=0.9\textheight,keepaspectratio]{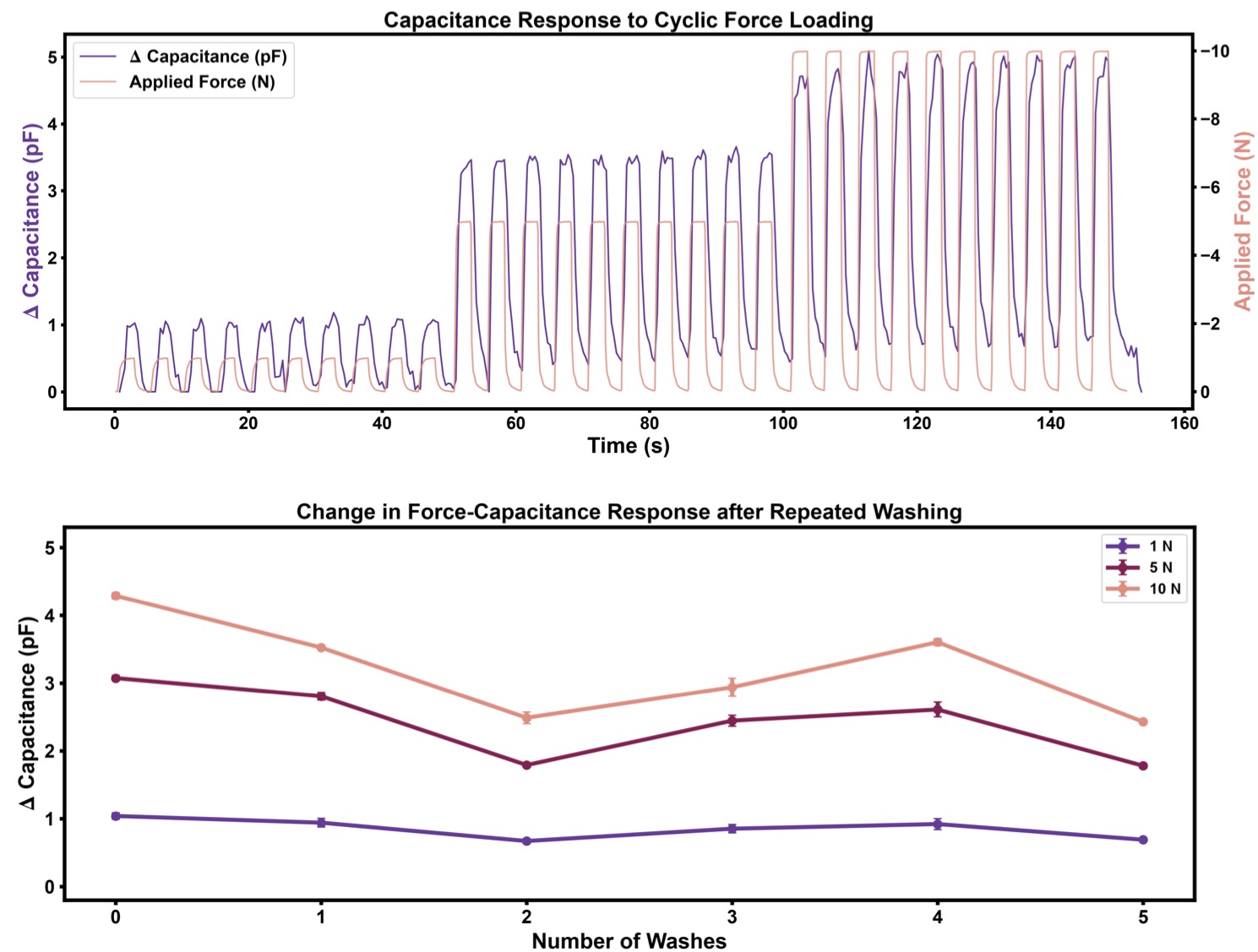}
    \caption{Force and capacitance change relationship for cyclically applied forces from 1N, 5N, and 10N (top) and subsequent changes after machine washing fabric for same cyclical force application (bottom).}
    \label{fig:forcewash}
\end{figure}

Despite the 2-layer row-column grid being a well studied method for capacitive tactile sensing \cite{grosse-puppendahl_finding_2017, fei_design_2024}, we introduce it's novel application directly in high-density knit structures. Our sensors currently have a density of 50 knitaxels/$cm^2$ and applications of touch localization and multi-touch distinguishability are shown in an 8 row x 8 column sensing patch in Figure \ref{fig:application} C. We conduct initial characterizations of force discernibility performance of our sensors by subjecting them to cyclical forces between 1N - 10N distributed across the entire sensor face. We average readings from the central 4 row x 4 column knitaxels to ensure stability away from sensor edges. We also further test washability of our sensors by repeating identical cyclical force testing between 5 sets of machine washing. Results from such force and washability tests are shown in Figure \ref{fig:forcewash}.

Overall, more tests are needed to clearly map a relationship between capacitance and force, however our preliminary results show promising stability between both repeated forces and washing cycles, indicating good structural integrity of yarns and minimal loss in pressure sensitivity between washes.

%Include in following future works section:
%triboelectric nanogenerator (TENG) tactile sensing
\section{Discussion and Future Works}

Future work will expand the interactive capabilities enabled by our functional knitted textiles and further optimize coating parameters for enhanced knittability. First, knitting coated yarns with increasingly higher density can improve sensor performance and support more expressive wearable interfaces, enabling gesture recognition, soft input surfaces, and more natural embodied interaction. Next, extending the dip-coating process to additional dielectric or conductive coating materials may unlock new sensing modalities for passive, non-invasive monitoring in health, athletics, and everyday activity. Moving towards novel coating materials that are both relevant for sensing and still mechanically compliant for knitting will be a large focus of future work. Some specific directions include experimenting with softer TPU variations (such as 60A or 70A) that may allow for similar insulation properties while further decreasing bending stiffness.

Furthermore, by pairing conductive cores with coatings engineered for strong thermoelectric or triboelectric responses, thermoelectric generators (TEGs) and triboelectric nanogenerators (TENGs) can be incorporated into knitted structures that harvest ambient energy to power low‑energy sensing, communication, or hybrid TENG–capacitive systems \cite{huo_recent_2025}. These energy‑aware systems move e‑textiles toward self‑powered, battery‑free platforms for long‑term wearable computing.

Finally, improving manufacturing uniformity and scalability of yarn fabrication will enrich development of textile devices with high spatial resolution and larger, complex form factor. For example, this enables enhanced, full-body robotic tactile perception and reduced reliance on vision-only pipelines or development of human wearables customized to each user and their desired application. The reliability of dip-coating uniformity and scalability of knitting as an additive manufacturing technique is essential for seamlessly integrating dense, durable sensing arrays within garments. Together, these directions position knitted sensing textiles as a promising platform for human-computer interaction and enable soft, continuous, and responsive interfaces to be integrated directly into the fabrics people wear and interact with everyday.

\bibliographystyle{ACM-Reference-Format}
\bibliography{TEMS_Bibliography}

\end{document}